# A flexible Bayesian generalized linear model for dichotomous response data with an application to text categorization

### Susana Eyheramendy[1] and David Madigan[2]

*Oxford University and Rutgers University*

**Abstract:** We present a class of sparse generalized linear models that include probit and logistic regression as special cases and offer some extra flexibility. We provide an EM algorithm for learning the parameters of these models from data. We apply our method in text classification and in simulated data and show that our method outperforms the logistic and probit models and also the elastic net, in general by a substantial margin.

## 1. Introduction

The standard approach to model the dependence of binary data on explanatory variables under the *Generalized linear models* setting, is through a cumulative density function (cdf) $\Psi$. For example, for a vector of explanatory variables $\boldsymbol{x}$ and a random variable $y$ that takes values in $\{0, 1\}$, the conditional probability of $y$ given $\boldsymbol{x}$ and a vector of parameters $\boldsymbol{\beta}$ is modelled using a cdf $\Psi$, i.e. $Pr(y = 1|\boldsymbol{x}, \boldsymbol{\beta}) = \Psi(\boldsymbol{x}^T\boldsymbol{\beta})$. The most commonly used cdfs are the logistic and normal cdfs. The corresponding "link functions", $\Psi^{-1}$, are the logit and probit link functions respectively.

Albert and Chib [1] proposed a Student-t inverse cdf as the link funtion. This model includes the logit and probit as special cases, at least approximately. For probabilities between 0.001 and 0.999, logistic quantiles are approximately a linear function of the quantiles of a Student-t distribution with 8 degrees of freedom. Also, when the degrees of freedom in a Student-t distribution are large, (say $v > 20$) the t-link approximates the probit model. The degrees of freedom $v$ control the thickness of the tail of the t-density allowing for a more flexible model. One can also benefit from this model as it can be presented via a latent variable representation that allows one to estimate the parameters of the model easily using the EM algorithm (Dempster et al. [5]).

For the logistic, normal and Student-t, the corresponding probability density functions (pdf) are symmetric around zero. This implies that their corresponding cdfs approach 1 at the same rate that they approach 0, which may not always be reasonable. In some applications, the overall fit can significantly improve by using a cdf that approaches 0 and 1 at different rates.

Many authors have proposed models with asymmetric link functions that can approximate, or have as special cases, the logit and probit links. Stukel [15] proposes a class of models that generalizes the logistic model. Chen et al. [4] introduce an alternative to the probit models where the cdf is the skew-normal distribution from Azzalini and Della Valle [2]. Fernandez and Steel [6] propose a class of skewed densities indexed by a scalar $\delta \in (0, \infty)$ of the form:

$$(1) \qquad p(y|\delta) = \frac{2}{\delta + \frac{1}{\delta}} \{ f(\frac{y}{\delta}) I_{[0,\infty)}(y) + f(y\delta) I_{(-\infty,0)}(y) \}, \qquad y \in \Re$$

where $f(.)$ is a univariate probability density function (pdf) with the mode at 0 and symmetry around the mode. The parameter $\delta$ determines the amount of mass at each side of the mode, and hence the skewness of the densities. Capobianco [3]) considers the Student-t pdf as the univariate $f(.)$ density in equation (1). This is an appealing model as it contains a parameter that controls the thickness of the tails and another parameter that determines the skewness of the density.

---

[1]Statistics Dept., Oxford University, Oxford, OX1 3TG, United Kingdom
[2]Department of Statistics, 501 Hill Center, Rutgers University, Piscataway, NJ 08855, USA, e-mail: madigan@stat.rutgers.edu

We apply an extended version of this model to textual datasets, where the problem is to classify text documents into predefined categories. We consider a Bayesian hierarchical model that contains, in addition to the parameter that controls the skewness of the density and the parameter that controls the thickness of the tails, a third parameter that controls the sparseness of the model, i.e. the number of regression parameters with zero posterior mode. In studies when there are a large number of predictor variables, this methodology gives one way of discriminating and selecting relevant predictors.

In what follows we describe the model that we propose.

## 2. The model

Suppose that $n$ independent binary random variables $Y_1, \ldots, Y_n$ are observed together with a vector of predictors $\mathbf{x}_1, \ldots, \mathbf{x}_n$, where $Y_i = 1$ with probability $P(Y_i = 1|\boldsymbol{\beta}, \mathbf{x}_i)$. Under the generalized linear model setting, models for binary classification satisfy $P(Y_i = 1|\boldsymbol{\beta}, \mathbf{x}_i) = \Psi(\mathbf{x}_i^T \boldsymbol{\beta})$ where $\Psi$ is a nonnegative function whose range is between 0 and 1. For instance, the probit model is obtained when $\Psi$ is the normal cumulative distribution and the logit model when $\Psi$ is the logistic cdf.

Under Bayesian learning one starts with a prior probability distribution for the unknown parameters of the model. Prediction of new data utilizes the posterior probability distribution. More specifically, denote by $\pi(\boldsymbol{\beta})$ the prior density function for the unknown parameter vector $\boldsymbol{\beta}$. Then the posterior density of $\boldsymbol{\beta}$ is given by

$$\pi(\boldsymbol{\beta}|\{(\mathbf{x}_1, y_1), \ldots, (\mathbf{x}_n, y_n)\}) = \frac{\pi(\boldsymbol{\beta}) \prod_{i=1}^{n} \Psi(\mathbf{x}_i^T \boldsymbol{\beta})^{y_i} (1 - \Psi(\mathbf{x}_i^T \boldsymbol{\beta}))^{1-y_i}}{\int \pi(\boldsymbol{\beta}) \prod_{i=1}^{n} \Psi(\mathbf{x}_i^T \boldsymbol{\beta})^{y_i} (1 - \Psi(\mathbf{x}_i^T \boldsymbol{\beta}))^{1-y_i} d\boldsymbol{\beta}}$$

and the posterior predictive distribution for $y$ given $\mathbf{x}$ is

$$\pi(y|\mathbf{x}, \{(\mathbf{x}_1, y_1), \ldots, (\mathbf{x}_n, y_n)\}) = \int Pr(y|\mathbf{x}, \boldsymbol{\beta}) \pi(\boldsymbol{\beta}|\{(\mathbf{x}_1, y_1), \ldots, (\mathbf{x}_n, y_n)\}) d\boldsymbol{\beta}$$

which in general are intractable due to the many integrals.

In the model that is proposed in this paper, we estimate the vector of parameters $\boldsymbol{\beta}$ as the mode of the posterior density $\pi(\boldsymbol{\beta}|\{(\mathbf{x}_1, y_1), \ldots, (\mathbf{x}_n, y_n)\})$ and prediction of new data is performed using the following rule:

$$(2) \qquad \hat{y} = 1 \text{ if } \pi(y = 1|\mathbf{x}, \{(\mathbf{x}_1, y_1), \ldots, (\mathbf{x}_n, y_n)\}) > 0.5,$$
$$\hat{y} = 0 \text{ if } \pi(y = 1|\mathbf{x}, \{(\mathbf{x}_1, y_1), \ldots, (\mathbf{x}_n, y_n)\}) \leq 0.5$$

which is an optimal rule in the sense that it has the smallest expected prediction error (see Hastie et al. [10] for more details).

We consider two general cases for the form of $\Psi$. First, we consider a cdf $\Psi$ that approaches 1 and 0 at the same rate. This is referred to hereon as the *symmetric case*. Second, we generalize this model assuming that the inverse of the link function is the cdf of a skewed density. In this way, the inverse of the link function approaches 0 and 1 at different rates.

### 2.1. Symmetric case

The model that we propose assumes that the conditional distribution of $Y_i$ given a vector of predictors $\boldsymbol{x}$ and a vector of unknown parameters $\boldsymbol{\beta}$ is determined by the cdf of the Student-t distribution with $v$ degrees of freedom evaluated at $\boldsymbol{x}_i^T \boldsymbol{\beta}$ i.e. $P(Y_i = 1|\boldsymbol{\beta}, \mathbf{x}_i) = T_v(\boldsymbol{x}_i^T \boldsymbol{\beta})$. Figure 1 shows how the Student-t cdf can approximate the normal and logistic cdfs. The black continuous line corresponds to the normal cdf, the dashed line correspond to the logistic cdf and the dotted lines to the Student-t cdf with different degrees of freedom. Also Figure 2 shows the logistic quantiles against the quantiles of a Student-t distribution with 8 degrees of freedom for probabilities between 0.0005 and 0.9995. The straight line corresponds to the linear model fit.

Assume that apriori the distribution of $\beta_j$ is normal with mean 0 and variance $\tau_j$, $N(0, \tau_j)$ and the distribution of the hyperparameters $\tau_j$ is exponential $exp(2/\gamma)$ with pdf equal to $\frac{\gamma}{2}e^{-\gamma\tau_j/2}$. Integrating with respect to $\tau_j$, one obtains $Pr(\beta_j|\gamma) = \frac{\sqrt{\gamma}}{2}exp(-\sqrt{\gamma}\|\beta_j\|)$, which is the double exponential prior. In Section 3 it will became clear that decomposing the double exponential into a two-level Bayesian hierarchical model, allows us to estimate the parameters $\boldsymbol{\beta}$ via the EM algorithm, where in addition to the latent variables $\mathbf{Z}$ (introduced below) the $\tau_j$ parameters are seen as missing data.

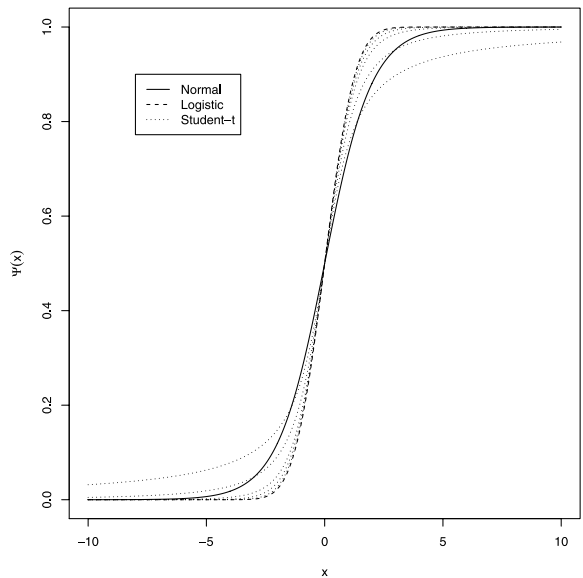

FIG 1. *The black continuous line corresponds to the normal cdf, the dashed line correspond to the logistic cdf and the dotted lines to the Student-t cdf with different degrees of freedom.*

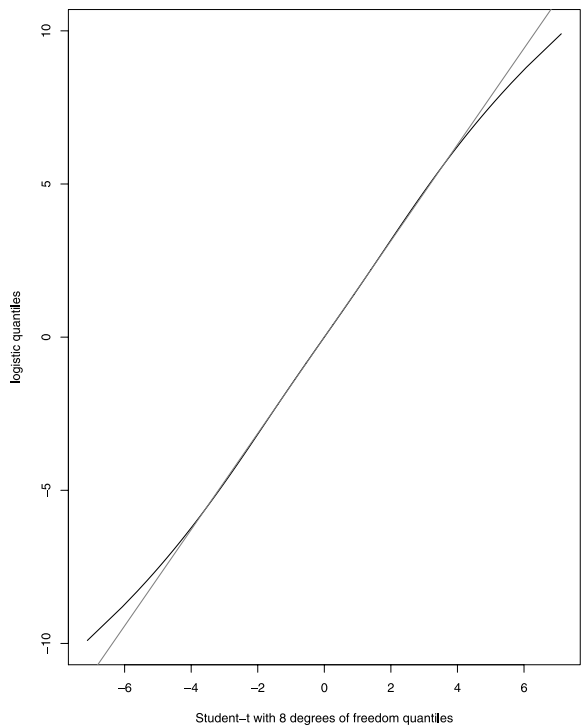

FIG 2. *Plot of the logistic quantiles against the quantiles of a Student-t distribution with 8 degrees of freedom for probabilities between 0.00005 and 0.99995. The strait line corresponds to the linear model fit.*

The normal prior on the $\boldsymbol{\beta}$ parameters has the effect of shrinking the maximun likelihood estimator of $\boldsymbol{\beta}$ towards zero, which has been shown to give a better generalization performance (see Hastie et al. [10] for more details). A different variance in the priors of the $\boldsymbol{\beta}$ gives the flexibility of having the parameters shrunk by a different amount which is relevant when the predictors influence the response unevenly. Moreover, the particular distribution of the variances $\tau_j$ that we use, will shrink some of the parameters $\beta_j$ to be exactly equal to zero. If the hyperprior distribution for $\tau_j$ places significant weight on small values of $\tau_j$ then it is likely that the estimate of the corresponding $\beta_j$ will also be small and can have zero posterior mode. On the other hand, hyperprior distributions that favor large values of the $\tau_j$'s will lead to posterior modes for the $\beta_j$'s that are close to the maximum likelihood estimates. Note that, since $E(\tau_j) = \frac{1}{\gamma}$ for all $j$, $\gamma$ effectively controls the sparseness of the model. Figure 3 shows the different shapes that the hyper distribution for $\tau_j$ can take as the parameter $\gamma$ varies. Analogous models have been applied by many authors e.g. Genkin et al [9], Figueiredo and Jain [7], Neal [13] and MacKay [11].

We show below that by introducing some latent variables, the generalized linear model that takes $\Psi$ to be equal to the Student-t cdf $T_v$ with $v$ degrees of freedom, can be seen as a missing data model and hence amenable to EM algorithm. This procedure offers a tractable way to estimate the parameters of the model. The latent variable representation for this model was first introduced by Albert and Chib [1], who derive a Gibbs sampling algorithm to find an estimator of the parameter $\boldsymbol{\beta}$. In the same spirit, Scott [14] introduced a latent variable representation for the logistic model.

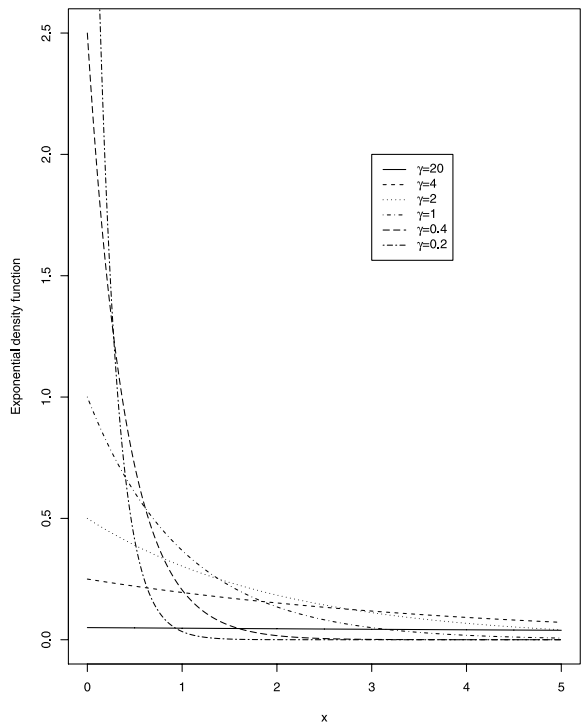

FIG 3. *Shapes that the exponential hyperprior for the $\tau_j$ parameters takes when the hyper parameter $\gamma$ changes.*

Let $Z_1, \ldots, Z_n$ be $n$ independent random variables where $Z_i$ is normally distributed with mean $\boldsymbol{x}_i^T \boldsymbol{\beta}$ and variance $\lambda_i^{-1}$, $N(\boldsymbol{x}_i^T \boldsymbol{\beta}, \lambda_i^{-1})$. Define the probability of $Y_i$ given $Z_i$ as a Bernoulli distribution with parameter equal to the probability that $Z_i$ is non-negative $Pr(Z_i > 0)$ i.e. $Y_i = 1$ if $Z_i > 0$ and $Y_i = 0$ otherwise. Let the inverse of the variance of $Z_i$, $\lambda_i$ be distributed as $Gamma(v/2, 2/v)$ with pdf proportional to $\lambda_i^{v/2-1} exp(-v\lambda_i/2)$.

Note that the marginal distribution $p(z|\boldsymbol{\beta}, \boldsymbol{x}, v)$ is $t_v(\boldsymbol{x}^T \boldsymbol{\beta})$, a Student-t density with $v$ degrees of freedom and location $\boldsymbol{x}^T \boldsymbol{\beta}$, and $p(y=1|\boldsymbol{\beta}, \boldsymbol{x}) = p(z > 0|\boldsymbol{\beta}, \boldsymbol{x}) = T_v(\boldsymbol{x}^T \boldsymbol{\beta})$ is the Student-t cumulative distribution with $v$ degrees of freedom, centered at 0 and evaluated at $\boldsymbol{x}^T \boldsymbol{\beta}$. Therefore, we see that by integrating the latent variables $z_i$ we recover the original model.

### 2.2. *General case*

Fernandez and Steel [6] propose a class of skewed densities indexed by a scalar $\delta \in (0, \infty)$ of the form:

$$p(y|\delta) = \frac{2}{\delta + \frac{1}{\delta}} \{f(\frac{y}{\delta}) I_{[0,\infty)}(y) + f(y\delta) I_{(-\infty,0)}(y)\}$$

where $f(.)$ is a univariate probability density function (pdf) with mode at 0 and symmetry around the mode. The parameter $\delta$ determines the amount of mass at each side of the mode, and hence the skewness of the densities. When $\delta$ varies between $(0, 1)$, the distribution has negative skewness and when $\delta$ is bigger than 1, the distribution has positive skewness. We replace the function $f()$ in equation (1) by the Student-t distribution and consider the corresponding cdf which we denote by $T_{v,\delta}$.

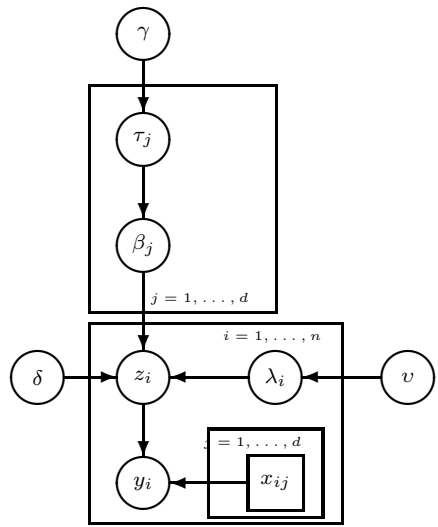

Fig 4. *Graphical model representation of the model with the skew Student-t link.*

Let $Y_1, \ldots, Y_n$ be $n$ binary independent random variables distributed Bernoulli with probability of success given by $P(Y_i = 1|\boldsymbol{\beta}, \mathbf{x}_i) = T_{v,\delta}(\boldsymbol{x}_i^T \boldsymbol{\beta})$. As in the symmetric case of Section 2.1, we introduce some latent variables that will allow us to find the parameter estimates using the EM algorithm.

Consider $n$ random variables $Z_1, \ldots, Z_n$ such that the distribution of $Z_i$ is the skew normal with mode at $\boldsymbol{x}_i^T \boldsymbol{\beta}$ given by

$$(3) \qquad p(z_i|\boldsymbol{x}_i, \boldsymbol{\beta}, \lambda_i, \delta) = \frac{2}{\delta + \frac{1}{\delta}} e^{-\frac{\lambda_i}{2}(r_i - \boldsymbol{x}_i^T \boldsymbol{\beta})^2} \sqrt{\frac{\lambda_i}{2\pi}}$$

where $r_i = \frac{z_i}{\delta} I(z_i \geq \boldsymbol{x}_i^T \boldsymbol{\beta}) + z_i \delta I(z_i < \boldsymbol{x}_i^T \boldsymbol{\beta})$. Define $Y_i = 1$ if $Z_i > 0$ and $Y_i = 0$ otherwise. Note that we can derive the pdf of equation (2) by replacing the $f()$ function of equation (1) with the normal distribution with mean $\boldsymbol{x}_i^T \boldsymbol{\beta}$ and variance $\lambda_i$ and accordingly divide the mass of the distribution at each side of the mode $\boldsymbol{x}_i^T \boldsymbol{\beta}$. Assume, as in the symmetric case, that *a priori* the distribution of $\beta_j$ is normal with mean 0 and variance $\tau_j$, the distribution of $\lambda_i$ is $Gamma(v/2, 2/v)$ and the distribution of the hyperparameters $\tau_j$ are exponential $exp(2/\gamma)$. See Figure 4 for a graphical model representation of this model.

Note that by marginalizing the pdf in equation (2) with respect to $\lambda_i$ we get,

$$p(z_i|\boldsymbol{x}_i, \boldsymbol{\beta}, \delta) = \frac{2}{\delta + \frac{1}{\delta}} \frac{\Gamma(\frac{v+1}{2})}{\Gamma(\frac{v}{2})\sqrt{\pi v}} \{1 + \frac{(z_i - \boldsymbol{x}_i^T \boldsymbol{\beta})^2}{v}$$

$$(4) \qquad \times [\frac{1}{\delta^2} I_{[\boldsymbol{x}_i^T \boldsymbol{\beta}, \infty)}(z_i) + \delta^2 I_{(-\infty, \boldsymbol{x}_i^T \boldsymbol{\beta})}(z_i)]\}^{-\frac{v+1}{2}}$$

which is the skew Student-t distribution with mode at $\boldsymbol{x}_i^T \boldsymbol{\beta}$, $v$ degrees of freedom and $\delta$ the parameter that controls the skewness of the pdf. Figure 5 shows this pdf for different vales of $\delta$ and 8 degrees of freedom. The continous line corresponds to the symmetric Student-t distribution. We can derive the pdf from equation (3) by replacing the function $f()$ from equation (1) with a Student-t distribution with mean $\boldsymbol{x}_i^T\boldsymbol{\beta}$ and $\upsilon$ degrees of freedom and accordingly divide the mass of the pdf at each side of the mode.

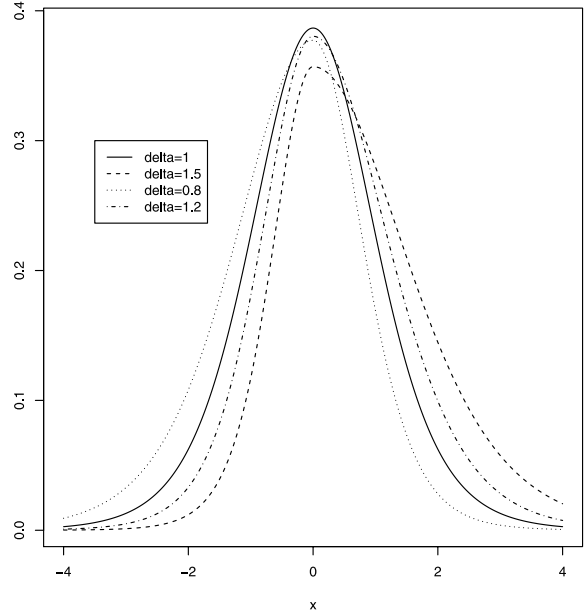

FIG 5. *This graph shows the different shapes that the skew Student-t distribution with 8 degrees of freedom. The continues line correspond to the symmetric case.*

## 3. The algorithm

### 3.1. The EM algorithm

In this section we derive the equations for the EM algorithm in the symmetric and general cases explained above.

The EM algorithm, originally introduced by Dempster, Laird and Rubin [5], provides an iterative procedure to compute *maximum likelihood estimators ($MLE$)*.

Each iteration in the EM algorithm has two steps, the E-step and the M-step.

E-step: Compute the expected value of the complete log-posterior, given the current estimate of the parameter and the observations, usually denoted as $Q$.

M-step: Find the parameter value that maximizes the $Q$ function.

### 3.2. Symmetric case

The complete log posterior for $\boldsymbol{\beta}$ is given by

$$
\begin{aligned}
\log p(\boldsymbol{\beta}|\mathbf{y},\mathbf{z},\lambda,\upsilon) &\propto -(\mathbf{z}-H\boldsymbol{\beta})^T A(\mathbf{z}-H\boldsymbol{\beta}) - \boldsymbol{\beta}^T W \boldsymbol{\beta} \\
&\propto 2\boldsymbol{\beta}^T H A \mathbf{z} - \boldsymbol{\beta}^T H^T A H \boldsymbol{\beta} - \boldsymbol{\beta}^T W \boldsymbol{\beta}
\end{aligned}
\tag{5}
$$

where $A = diag(\lambda_1, \ldots, \lambda_n)$, $H = [\mathbf{x_1}, \ldots, \mathbf{x_n}]^T$ is the design matrix and $W = diag(\tau_1^{-1}, \ldots, \tau_m^{-1})$ is the covariance matrix for the normal prior on $\boldsymbol{\beta}$.

From equation (5) we see that the matrices $A$ and $W$ and the vector $\mathbf{z}$ correspond to the missing data. Therefore, the E-step needs to compute $E(\tau_i^{-1}|\mathbf{y}, \hat{\boldsymbol{\beta}}^{(t)}, \gamma, v)$, $E(\lambda_i|\mathbf{y}, \hat{\boldsymbol{\beta}}^{(t)}, \gamma, v)$ and $E(z_i\lambda_i|\mathbf{y}, \hat{\boldsymbol{\beta}}^{(t)}, \gamma, v)$ for $i = 1, \ldots, n$.

In order to do these computations we need first to get the pdfs $p(z_i|\hat{\boldsymbol{\beta}}^{(t)}, y_i, \lambda_i, \gamma, v)$ and $p(\lambda_i|\hat{\boldsymbol{\beta}}^{(t)}, y_i, v)$. Note that the conditional probability of $\mathbf{z}_i$ given $\hat{\boldsymbol{\beta}}^{(t)}$, $y_i$, $\lambda_i$, $\gamma$ and $v$ is a normal distribution with mean $\mathbf{x}_i^T\beta$ and variance $\lambda_i$ but left truncated at zero if $y_i = 1$ and right truncated at zero if $y_i = 0$. The posterior probabilities for $\lambda_i$ also have closed form given by,

$$p(\lambda_i|\hat{\boldsymbol{\beta}}^{(t)}, y, v) = \begin{cases} \frac{p(\lambda_i|v)\Phi(\sqrt{\lambda_i}\boldsymbol{x}_i^T\boldsymbol{\beta})}{T_v(\boldsymbol{x}_i^T\boldsymbol{\beta})} & y_i = 1, \\ \\ \frac{p(\lambda_i|v)\Phi(-\sqrt{\lambda_i}\boldsymbol{x}_i^T\boldsymbol{\beta})}{T_v(-\boldsymbol{x}_i^T\boldsymbol{\beta})} & y_i = 0, \end{cases}$$

where $\Phi$ denotes the cumulative distribution for the standard normal with mean 0 and variance 1.

Now we can compute the expected value for $\lambda_i$ and $z_i$, which are given by

$$(6) \qquad E(\lambda_i|\hat{\boldsymbol{\beta}}^{(t)}, y_i, v) = \begin{cases} \frac{T_{v+2}(\boldsymbol{x}_i^T\boldsymbol{\beta}\sqrt{\frac{v+2}{v}})}{T_v(\boldsymbol{x}_i^T\boldsymbol{\beta})} & y_i = 1, \\ \\ \frac{T_{v+2}(-x_i^T\boldsymbol{\beta}\sqrt{\frac{v+2}{v}})}{T_v(-\boldsymbol{x}_i^T\boldsymbol{\beta})} & y_i = 0, \end{cases}$$

and

$$(7) \qquad \begin{aligned} &E(z_i|\hat{\beta}^{(t)}, \lambda_i, y, v) = \\ &\begin{cases} \hat{\boldsymbol{\beta}}^{(t)}h(\mathbf{x^{(i)}}) + \frac{1}{\sqrt{\lambda_i}}\frac{\phi(\sqrt{\lambda_i}\hat{\beta}^{(t)}h(\mathbf{x^{(i)}}))}{\Phi(\sqrt{\lambda_i}\hat{\beta}^{(t)}h(\mathbf{x^{(i)}}))} & y_i = 1, \\ \\ \hat{\boldsymbol{\beta}}^{(t)}h(\mathbf{x^{(i)}}) - \frac{1}{\sqrt{\lambda_i}}\frac{\phi(\sqrt{\lambda_i}\hat{\beta}^{(t)}h(\mathbf{x^{(i)}}))}{\Phi(-\sqrt{\lambda_i}\hat{\beta}^{(t)}h(\mathbf{x^{(i)}}))} & y_i = 0, \end{cases} \end{aligned}$$

respectively. The next computation in the E-step is $E(\lambda_iz_i|y_i, \hat{\boldsymbol{\beta}}^{(t)}, v)$. Observe that $E(\lambda_iz_i|y_i, \hat{\boldsymbol{\beta}}^{(t)}, v) = E\{E(\lambda_iz_i|y_i, \hat{\boldsymbol{\beta}}^{(t)}, v, \lambda_i)\} = E\{\lambda_iE(z_i|y_i, \hat{\boldsymbol{\beta}}^{(t)}, v, \lambda_i)\}$. Therefore by replacing first eq. (7) and then eq. (6) in this expectation we obtain,

$$(8) \qquad \begin{aligned} &E(\lambda_iz_i|y_i, \hat{\boldsymbol{\beta}}^{(t)}, v) = \\ &\begin{cases} \frac{\boldsymbol{x}^T\boldsymbol{\beta}T_{v+2}(\boldsymbol{x}^T\boldsymbol{\beta}\sqrt{\frac{v+2}{v}})}{T_v(\boldsymbol{x}^T\boldsymbol{\beta})} + \frac{t_v(\boldsymbol{x}^T\boldsymbol{\beta})}{T_v(\boldsymbol{x}^T\boldsymbol{\beta})} & y_i = 1, \\ \\ \frac{\boldsymbol{x}^T\boldsymbol{\beta}T_{v+2}(-\boldsymbol{x}^T\boldsymbol{\beta}\sqrt{\frac{v+2}{v}})}{T_v(-\boldsymbol{x}^T\boldsymbol{\beta})} - \frac{t_v(\boldsymbol{x}^T\boldsymbol{\beta})}{T_v(-\boldsymbol{x}^T\boldsymbol{\beta})} & y_i = 0. \end{cases} \end{aligned}$$

Finally, the expectation for $\tau_j$ is given by,

$$(9) \qquad \begin{aligned} E(\tau_j^{-1}|\mathbf{y}, \hat{\boldsymbol{\beta}}^{(t)}, \gamma, v) &= \frac{\int_0^\infty \frac{1}{\tau_j}p(\tau_j|\gamma)p(\hat{\boldsymbol{\beta}}^{(t)}|\tau_j)d\tau_j}{\int_0^\infty p(\tau_j|\gamma)p(\hat{\boldsymbol{\beta}}^{(t)}|\tau_j)d\tau_j} \\ &= \frac{\gamma}{|\beta_j|}. \end{aligned}$$

Denote,

$$W^* = diag(E(\tau_1^{-1}|\mathbf{y}, \hat{\boldsymbol{\beta}}^{(t)}, \gamma, v), \ldots, E(\tau_m^{-1}|\mathbf{y}, \hat{\boldsymbol{\beta}}^{(t)}, \gamma, v)),$$
$$A^* = diag(E(\lambda_1|\mathbf{y}, \hat{\boldsymbol{\beta}}^{(t)}, \gamma, v), \ldots, E(\lambda_n|\mathbf{y}, \hat{\boldsymbol{\beta}}^{(t)}, \gamma, v)),$$
$$\mathbf{z}^* = (E(z_1\lambda_1|\mathbf{y}, \hat{\boldsymbol{\beta}}^{(t)}, \gamma, v), \ldots, E(z_n\lambda_n|\mathbf{y}, \hat{\boldsymbol{\beta}}^{(t)}, \gamma, v))^T.$$

Then the M-step that results from maximizing equation (5) with respect to $\boldsymbol{\beta}$ is given by,

$$(10) \qquad \hat{\beta} = (H^T A^* H + W^*)^{-1} H^T \mathbf{z}^*$$

Summarizing, in the $(t+1)$th iteration of the EM algorithm,

**E-step**: Compute $W^*$, $A^*$ and $\mathbf{z}^*$ using equations (8), (5) and (7) respectively.

**M-step**: Obtain a new estimate $\hat{\boldsymbol{\beta}}^{(t+1)}$ by replacing the new values of $W^*$, $A^*$ and $\mathbf{z}^*$ in eq. (9). If $\|\hat{\boldsymbol{\beta}}^{(t+1)} - \hat{\boldsymbol{\beta}}^{(t)}\|/\|\hat{\boldsymbol{\beta}}^{(t)}\| < \Delta$ stop, otherwise go back to the E-step. We fix $\Delta = 0.005$ following Figuereido and Jain [7].

### 3.3. General case

In the general case of Section 4.2, the complete log-posterior can be written as

$$
\begin{aligned}
\log p(\boldsymbol{\beta}|\lambda, v, \mathbf{y}, \mathbf{z}, \delta) \\
\propto -(\mathbf{r} - H\boldsymbol{\beta})^T A(\mathbf{r} - H\boldsymbol{\beta}) - \boldsymbol{\beta}^T W \boldsymbol{\beta} \\
\propto 2\boldsymbol{\beta}^T H A \mathbf{r} - \boldsymbol{\beta}^T H^T A H \boldsymbol{\beta} - \boldsymbol{\beta}^T W \boldsymbol{\beta}
\end{aligned}
\tag{11}
$$

where $r_i = \frac{z_i}{\delta}I(z_i > 0) + z_i\delta I(z_i < 0)$, $W$ and $A$ are defined in Section 5.1. To get the new equations for the EM algorithm we have to compute $E(\lambda_i r_i|y_i, \hat{\beta}^{(t)}, v)$ and $E(\lambda_i|y_i, \hat{\boldsymbol{\beta}}^{(t)}, v)$ for $i = 1, \ldots, n$. Using the same trick that we used before, i.e. $E(\lambda_i z_i|y_i, \hat{\boldsymbol{\beta}}^{(t)}, v, \delta) = E\{E(\lambda_i z_i|y_i, \hat{\boldsymbol{\beta}}^{(t)}, v, \lambda_i, \delta)\} = E\{\lambda_i E(z_i|y_i, \hat{\boldsymbol{\beta}}^{(t)}, v, \lambda_i, \delta)\}$, we obtain $E(\lambda_i r_i|y_i, \hat{\boldsymbol{\beta}}^{(t)}, v, \lambda_i)$ for $y_i = 0$ and $y_i = 1$ in the following way:
$E(\lambda_i r_i|y_i = 1, \boldsymbol{\beta}, v, \lambda_i) =$

$$(12) \quad
\begin{cases}
\frac{\delta \boldsymbol{x}_i^T \boldsymbol{\beta} T_{v+2,\delta}(-\boldsymbol{x}_i^T \boldsymbol{\beta}(\frac{1}{\delta}-1)\sqrt{\frac{v+2}{v}})}{T_{v,\delta}(\boldsymbol{x}_i^T \boldsymbol{\beta})} + \delta \frac{t_{v,\delta}(\boldsymbol{x}_i^T \boldsymbol{\beta}(\frac{1}{\delta}-1))}{T_{v,\delta}(\boldsymbol{x}_i^T \boldsymbol{\beta})} \\
+ \frac{\boldsymbol{x}_i^T \boldsymbol{\beta}}{\delta} \frac{\{T_{v+2,\delta}(\boldsymbol{x}_i^T \boldsymbol{\beta}(\delta-1)\sqrt{\frac{v+2}{v}}) - T_{v,\delta}(-\boldsymbol{x}_i^T \boldsymbol{\beta}\sqrt{\frac{v+2}{v}})\}}{T_{v,\delta}(\boldsymbol{x}_i b^T \boldsymbol{\beta})} \\
+ \frac{1}{\delta} \frac{\{t_{v,\delta}(-\boldsymbol{x}_i^T \boldsymbol{\beta}) - t_{v,\delta}(\boldsymbol{x}_i^T \boldsymbol{\beta}(\delta-1))\}}{T_{v,\delta}(\boldsymbol{x}_i^T \boldsymbol{\beta})} & x_i^T \boldsymbol{\beta} \geq 0, \\
\frac{\delta \boldsymbol{x}_i^T \boldsymbol{\beta} T_{v+2,\delta}(\boldsymbol{x}_i^T \boldsymbol{\beta}\sqrt{\frac{v+2}{v}})}{T_{v,}(\boldsymbol{x}_i^T \boldsymbol{\beta})} + \delta \frac{t_{v,\delta}(-\boldsymbol{x}_i^T \boldsymbol{\beta})}{T_{v,\delta}(\boldsymbol{x}_i^T \boldsymbol{\beta})} & \boldsymbol{x}_i^T \boldsymbol{\beta} < 0,
\end{cases}
$$

$E(\lambda_i r_i|y_i = 0, \boldsymbol{\beta}, v, \lambda_i) =$

$$(13) \quad
\begin{cases}
\frac{\boldsymbol{x}_i^T \boldsymbol{\beta}}{\delta} \frac{T_{v+2,\delta}(-\boldsymbol{x}_i^T \boldsymbol{\beta}\sqrt{\frac{v+2}{v}})}{T_{v,\delta}(-\boldsymbol{x}_i^T \boldsymbol{\beta})} - \frac{1}{\delta} \frac{t_{v,\delta}(\boldsymbol{x}_i^T \boldsymbol{\beta})}{T_{v,\delta}(-\boldsymbol{x}_i^T \boldsymbol{\beta})} & \boldsymbol{x}_i^T \boldsymbol{\beta} \geq 0, \\
\delta \boldsymbol{x}_i^T \boldsymbol{\beta} \frac{T_{v+2,\delta}(-\boldsymbol{x}_i^T \boldsymbol{\beta}\sqrt{\frac{v+2}{v}}) - T_{v+2,\delta}(\boldsymbol{x}_i^T \boldsymbol{\beta}(\frac{1}{\delta}-1)\sqrt{\frac{v+2}{v}})}{T_{v,\delta}(-\boldsymbol{x}_i^T \boldsymbol{\beta})} \\
+ \delta \frac{t_{v,\delta}(\boldsymbol{x}_i^T \boldsymbol{\beta}(\frac{1}{\delta}-1)) - t_{v,\delta}(-\boldsymbol{x}_i^T \boldsymbol{\beta})}{T_{v,\delta}(-\boldsymbol{x}_i^T \boldsymbol{\beta})} - \frac{1}{\delta} \frac{t_{v,\delta}(\boldsymbol{x}_i^T \boldsymbol{\beta}(\delta-1))}{T_{v,\delta}(-\boldsymbol{x}_i^T \boldsymbol{\beta})} \\
+ \frac{\boldsymbol{x}_i^T \boldsymbol{\beta}}{\delta} \frac{T_{v+2,\delta}(\boldsymbol{x}_i^T \boldsymbol{\beta}(\delta-1)\sqrt{\frac{v+2}{v}})}{T_{v,\delta}(-\boldsymbol{x}_i^T \boldsymbol{\beta})} & \boldsymbol{x}_i^T \boldsymbol{\beta} < 0.
\end{cases}
$$

The expected value for $\lambda_i$ given $y_i, \hat{\boldsymbol{\beta}}^{(t)}, v$, $E(\lambda_i|y_i,\hat{\boldsymbol{\beta}}^{(t)}, v)$, can be computed as follows:

$$(14) \quad E(\lambda_i|y_i=1,\boldsymbol{\beta},v) = \begin{cases} \frac{1}{\delta}\frac{T_{v+2,\delta}(\boldsymbol{x}_i^T\boldsymbol{\beta}(\delta-1)\sqrt{\frac{v+2}{v}})}{T_{v,\delta}(\boldsymbol{x}_i^T\boldsymbol{\beta})} - \\ \frac{1}{\delta}\frac{T_{v+2,\delta}(-\boldsymbol{x}_i^T\boldsymbol{\beta}\sqrt{\frac{v+2}{v}})}{T_{v,\delta}(\boldsymbol{x}_i^T\boldsymbol{\beta})} + \\ \delta\frac{T_{v+2,\delta}(-\boldsymbol{x}_i^T\boldsymbol{\beta}(\frac{1}{\delta}-1)\sqrt{\frac{v+2}{v}})}{T_{v,\delta}(\boldsymbol{x}_i^T\boldsymbol{\beta})} & \boldsymbol{x}_i^T\boldsymbol{\beta}\geq 0, \\ \delta\frac{T_{v+2,\delta}(\boldsymbol{x}_i^T\boldsymbol{\beta}\sqrt{\frac{v+2}{v}})}{T_{v,\delta}(\boldsymbol{x}_i^T\boldsymbol{\beta})} & \boldsymbol{x}_i^T\boldsymbol{\beta}< 0, \end{cases}$$

$$(15) \quad E(\lambda_i|y_i=0,\boldsymbol{\beta},v) = \begin{cases} \frac{1}{\delta}\frac{T_{v+2,\delta}(-\boldsymbol{x}_i^T\boldsymbol{\beta}\sqrt{\frac{v+2}{v}})}{T_{v,\delta}(-\boldsymbol{x}_i^T\boldsymbol{\beta})} & \boldsymbol{x}_i^T\boldsymbol{\beta}\geq 0, \\ \frac{1}{\delta}\frac{T_{v+2,\delta}(\boldsymbol{x}_i^T\boldsymbol{\beta}(\delta-1)\sqrt{\frac{v+2}{v}})}{T_{v,\delta}(-\boldsymbol{x}_i^T\boldsymbol{\beta})} + \\ \delta\frac{T_{v+2,\delta}(-\boldsymbol{x}_i^T\boldsymbol{\beta}\sqrt{\frac{v+2}{v}})}{T_{v,\delta}(-\boldsymbol{x}_i^T\boldsymbol{\beta})} - \\ \delta\frac{T_{v+2,\delta}(\boldsymbol{x}_i^T\boldsymbol{\beta}(\frac{1}{\delta}-1)\sqrt{\frac{v+2}{v}})}{T_{v,\delta}(-\boldsymbol{x}_i^T\boldsymbol{\beta})} & \boldsymbol{x}_i^T\boldsymbol{\beta}< 0. \end{cases}$$

Denote,

$$(16) \quad r^* = (E(\lambda_1 r_1),\ldots,E(\lambda_n r_n)),$$

$$(17) \quad W^* = diag(E(\tau_1^{-1}|\mathbf{y},\hat{\boldsymbol{\beta}}^{(t)},\gamma,v),\ldots,E(\tau_m^{-1}|\mathbf{y},\hat{\boldsymbol{\beta}}^{(t)},\gamma,v)),$$

$$(18) \quad A^* = diag(E(\lambda_1|\mathbf{y},\hat{\boldsymbol{\beta}}^{(t)},\gamma,v),\ldots,E(\lambda_n|\mathbf{y},\hat{\boldsymbol{\beta}}^{(t)},\gamma,v)).$$

The new steps for the EM algorithm are as follows. In the $(t+1)$th iteration,

**E-step**: Compute $W^*, A^*$ and $\mathbf{r}^*$ using equations $(8), (12)$ and $(11)$ respectively.

**M-step**: Obtain a new estimate $\hat{\boldsymbol{\beta}}^{(t+1)}$ by replacing the new values of $W^*, A^*$ and $\mathbf{r}^*$ in eq. $(9)$, where now $z^*$ is replaced by $r^*$. If $\|\hat{\boldsymbol{\beta}}^{(t+1)} - \hat{\boldsymbol{\beta}}^{(t)}\|/\|\hat{\boldsymbol{\beta}}^{(t)}\| < \Delta$ stop, otherwise go back to the E-step. We set $\Delta = 0.005$.

In both cases, the symmetric and general, we initialized the algorithm by setting $\hat{\boldsymbol{\beta}}^{(0)} = (\epsilon I + H^T H)^{-1} H\mathbf{y}$ (with $\epsilon = 1e-6$), which corresponds to a weakly penalized ridge-type estimator.

## 4. Simulation study

In this section we present a simulation study that compares the flexible Student-t link (hereafter FST) with the probit and logistic models and also with the recently introduced "elasticnet" (Zou and Hastie [16]).

We assess the performance of the models by measuring the misclassification rate using simulated datasets. Two examples are presented here.

Example 1: We generated 10 datasets consisting of 10 predictors and 100 observations. The response random variable was generated as a random binomial of size 1 and probability 0.6 and the 10 predictors were generated as a random binomial of size 1 and probabilities equal to 0.3 (in three predictors), 0.5 (in five predictors) and 0.8 (in two predictors).

Example 2: We generated 10 datasets consisting of 10 predictors and 100 observations. We allow in these datasets high correlation between the response variable and the predictors, and also high correlation within the predictors. We generated the response and the predictors as multivariate normal with mean zero and some structure in the covariate matrix and then we dichotomized these variables by assigning a 0 to negative values and 1 to positive values. The assumed covariance matrix structure is the following: the response together with the first 4 predictors all have correlation equal to 0.8 with each other. The next three predictors have correlation 0.3 with each other, and the last three predictors have correlation 0.4 with each other. The other correlations are all equal to 0.01.

Our model has three tuning parameters, $v$ (the degrees of freedom of the Student-t distribution) that controls the thickness of the tails of the distribution, $\gamma$ that controls the sparseness of the parameter estimates and $\delta$ that controls the skewness of the distribution. For different values of the parameters ($v \in \{1, \ldots, 8, 15, 30, 50, 100\}$, $\gamma \in \{0.01, 0.1, 1, \ldots, 10, 20, 50, 100\}$ and $\delta \in \{0.01, 0.1, 0.5, 0.7, 1, 1.2, 1.5, 2, 3, 4, 5, 10\}$) we estimate the misclassification rate of our proposed model. We compare the best performance with the performance of the generalized linear models with probit and logit links which were fitted using the $R$ statistical package (see results in Table 1). In Example 1, our flexible $t$ link function FST consistently gives better performance than the logit and probit, ranging from 1% improvement in dataset 9 to 16% in the first dataset. Note that most of the best models choose the degrees of freedom of the Student-$t$ distribution to be equal to 1 i.e. they prefer fat-tailed distributions.

We also compare our FST model with a related method introduced by (Zou and Hastie (2005)), the so-called elastic net. We estimated this model over a grid of points for the two parameters of the model ($\lambda \in \{0.01, 0.1, 1, 10, 100\}$) using the statistical software R and selected the best performance. The results are shown in the last column of Table 1. Our method gives slightly poorer performance only in Datasets $9 - 10$.

We look at the best performance of the flexible probit (approximated by Student-$t_{30}$) and the logit (approximated by Student-$t_8$) models (results in Table 2) and their choice of the parameters that give best performance. Note that in most cases the same set of parameters gives best performance in both models. The three link functions consistently choose $\delta = 1.2$ in most datasets, i.e. they prefer skewed distributions.

Table 3 shows the results for the simulated datasets in Example 2. FST outperforms the three other methods in general by a substantial margin. We compare the sparseness of the elasticnet with the FST. In general, the choice of parameters for the best performed FST model, favors small values of the $\gamma$ parameter, which do not induced sparseness in the model. Compared to the elasticnet, the FST model appears to be less sparse. Results are shown in Table 4 for the 10 datasets of Example 2.

## 5. Applications to text categorization

Text categorization concerns the automated classification of documents into categories from a predefined set $C = \{c_1, \ldots, c_m\}$. A standard approach is to build $m$ independent binary classifiers, one for each category, that decide whether or not a document $d_j$ is in category $c_i$ for $i \in \{1, \ldots, m\}$ and $j \in \{1, \ldots, n\}$. Construction of the binary classifiers requires the availability of a corpus of documents $D$ with assigned categories. We learn the classifiers using a subset of the corpus $Tr \subset D$, the training set, and we test the classifiers using what remains of the corpus $Te = D/Tr$, the testing set.

TABLE 1

*Example 1. Misclassification rates for the generalized linear model with probit and logit link functions computed using the statistical software R are shown in the first two columns. The third column correspond to the best performance achieved, and the following three columns correspond to the parameters of the model that achieved the best performance. A ∗ in a cell means that the minimum misclassification rate is achieved by more than one value of the parameter.*

| Data | Misclassification | | | Parameters | | | |
|------|--------|-------|------|-----|-----|------|------|
|      | probit | logit | FST  | $v$ | $\delta$ | $\gamma$ | enet |
| 1    | 0.4    | 0.4   | 0.26 | 1   | ∗   | ∗    | 0.27 |
| 2    | 0.27   | 0.28  | 0.23 | 1   | 1.2 | 0.01 | 0.27 |
| 3    | 0.4    | 0.4   | 0.36 | ∗   | 1.2 | 0.1  | 0.36 |
| 4    | 0.4    | 0.4   | 0.38 | 1   | ∗   | 0.1  | 0.38 |
| 5    | 0.28   | 0.28  | 0.25 | 1   | ∗   | ∗    | 0.29 |
| 6    | 0.4    | 0.4   | 0.36 | 1   | ∗   | 0.01 | 0.38 |
| 7    | 0.34   | 0.33  | 0.30 | 1   | 1.2 | ∗    | 0.31 |
| 8    | 0.39   | 0.37  | 0.31 | ∗   | ∗   | ∗    | 0.36 |
| 9    | 0.32   | 0.32  | 0.31 | ∗   | ∗   | ∗    | 0.29 |
| 10   | 0.33   | 0.33  | 0.31 | ∗   | ∗   | ∗    | 0.28 |

TABLE 2

*Example 1. Best performance achieved by the probit and logit models when approximated by a $t_{30}$ and $t_8$ respectively with the correspondent parameters. A ∗ in a cell means that the minimum misclassification rate is achieved by more than one value of the parameter.*

| Data | Probit | Parameters | | Logit | Parameters | |
|------|--------|----------|----------|-------|----------|----------|
|      | probit | $\delta$ | $\gamma$ | logit | $\delta$ | $\gamma$ |
| 1    | 0.28   | 1.2      | 1        | 0.27  | 1.2      | 1        |
| 2    | 0.26   | ∗        | 0.1      | 0.26  | ∗        | 0.1      |
| 3    | 0.37   | 1.2      | 0.1      | 0.36  | 1.2      | 0.1      |
| 4    | 0.4    | ∗        | 1        | 0.4   | ∗        | 1        |
| 5    | 0.28   | ∗        | ∗        | 0.27  | 1.2      | 1        |
| 6    | 0.39   | ∗        | 0.1      | 0.39  | ∗        | 0.1      |
| 7    | 0.31   | 1.2      | 0.1      | 0.32  | 1.2      | 0.1      |
| 8    | 0.31   | ∗        | 3        | 0.32  | ∗        | 3        |
| 9    | 0.31   | ∗        | 1        | 0.31  | ∗        | 1        |
| 10   | 0.31   | ∗        | 1        | 0.31  | ∗        | 1        |

TABLE 3

*Example 2. Misclassification rates for the generalized linear model with probit and logit link functions computed using the statistical software R are shown in the first two columns. The third column correspond to the best performance achieved, and the following three columns correspond to the parameters of the model that achieved the best performance. A ∗ in a cell means that the minimum misclassification rate is achieved by more than one value of the parameter.*

| Data | Misclassification | | | Parameters | | | |
|------|--------|-------|------|-----|-----|------|------|
|      | probit | logit | FST  | $v$ | $\delta$ | $\gamma$ | enet |
| 1    | 0.08   | 0.08  | 0.03 | 1   | ∗   | ∗    | 0.06 |
| 2    | 0.15   | 0.14  | 0.06 | 1   | ∗   | 0.01 | 0.13 |
| 3    | 0.13   | 0.12  | 0.06 | 1   | ∗   | ∗    | 0.12 |
| 4    | 0.13   | 0.13  | 0.1  | ∗   | ∗   | ∗    | 0.11 |
| 5    | 0.08   | 0.08  | 0.05 | 1   | ∗   | ∗    | 0.08 |
| 6    | 0.11   | 0.11  | 0.08 | ∗   | ∗   | ∗    | 0.10 |
| 7    | 0.09   | 0.09  | 0.07 | ∗   | ∗   | ∗    | 0.09 |
| 8    | 0.13   | 0.14  | 0.09 | ∗   | ∗   | ∗    | 0.13 |
| 9    | 0.15   | 0.15  | 0.07 | 1   | ∗   | 0.01 | 0.14 |
| 10   | 0.18   | 0.18  | 0.12 | ∗   | ∗   | ∗    | 0.16 |

TABLE 4

*Number of parameters estimates equal to zero in the elasticnet and FST model for each of the datasets in Example 2.*

| model | 1 | 2 | 3 | 4 | 5 | 6 | 7 | 8 | 9 | 10 |
|-------|---|---|---|---|---|---|---|---|---|----|
| enet  | 0 | 0 | 0 | 2 | 5 | 3 | 9 | 6 | 3 | 7  |
| FST   | 1 | 0 | 0 | 0 | 2 | 1 | 4 | 0 | 0 | 5  |

Usually, documents are represented as vectors of weights $d_j = (x_{1j}, \ldots, x_{dj})$ where $x_{ij}$ represents a function of the frequency of appearance of word $w_i$ in document $d_j$ for $d$ words $w_1, \ldots, w_d$ in the "bag of words". This is the so-called "bag of words" representation (see e.g. Mladenic [12]).

Due to the large number of possible features or different words that can be gathered from a set of documents (usually this could be one hundred thousand or more) the classifiers are commonly built with a subset of the words. The text classification literature has tended to focus on feature selection (or word selection) algorithms that compute a score independently for each candidate feature, this is the so-called filtering approach. The scores typically depend on the counts of occurrences of the word in documents within the class and outside the class in training documents. For a predefined number of words to be selected, say $d$, the $d$ words with the highest score are chosen. Several score functions exist, for a thorough comparative study of many of them see Forman [8]. We consider the 100 best words, according to the information gain criterion. Before we selected these 100 words we remove common noninformative words taken from a standard *stopword* list of 571 words.

We performed the experiments in one standard text dataset. The dataset that we use comes from the Reuters news story collection that contains $21,450$ documents that have assigned zero or more categories to them among more than a hundred categories. We use a subset of the ModApte version of the Reuters$-21,578$ collection, where each document has assigned at least one topic label (or category) and this topic label belongs to any of the 10 most populous categories—earn, acq, grain, wheat, crude, trade, interest, corn, ship, money-fx. It contains $6,775$ documents in the training set and $2,258$ in the testing set.

To evaluate the performance of the different classifiers we use Recall, Precision and the $F_1$ measure. Recall measures the proportion of documents correctly classified within documents in the same category. Precision measures the proportion of documents correctly classified within all documents classified into the same category, and $F_1$ is the harmonic mean of Recall and Precision. There are two ways to average Recall, Precision and the $F_1$ measure over all categories, micro-averaged and macro-averaged. The micro-averaged is an average weighted by the class distribution and the macro-averaged is the arithmetic mean over all categories. All three measures depend on a specific threshold which is chosen by either doing cross validation or by letting part of the dataset (a validation set) determine the best choice, or by fixing it arbitrarily. We set this threshold to be 0.5, i.e., we simply classify a document to the category with the highest probability.

To choose the model that will perform best in new data, we divide the dataset into three parts: a training set, a validation set and a testing set. In the training set, we fix the tuning parameters and learn the model which is tested using the validation set. We repeat this process for every set of tuning parameters. We pick the set of tuning parameters that gives the best performance in the validation set. Then the algorithm learns the model with the chosen tuning parameters, this time using training and validation sets. The performance of the this final model is asessed using the testing set. We repeat this whole process 5 times, for different splits of the dataset into training-validation-testing sets to evaluate the performance error. We utilize 50% of the documents for training, 25% for validation and 25% for testing in the 5 splits of the dataset.

We vary the tuning parameters as follows: $\upsilon \in \{1, 2, 5, 8, 30\}$, $\gamma \in \{0.01, 0.05, 0.1, 1, 2\}$ and $\delta \in \{0.1, 0.5, 1, 1.5, 2\}$. For each category, we pick the set of tuning parameters $(\upsilon, \gamma, \delta)$ that gives the highest performance according to the $F_1$-measure in the validation test. Table 5 shows the values of these parameters for Split 1. Note that three of the categories (among the most populous ones) choose a symmetric link ($\delta = 1$).

The first column of Tables 6, 7 and 8 shows the micro and macro average of the $F_1$ measure, recall and precision respectively, of the performance of the FST models in the testing set for the 5 splits of the dataset.

The second column corresponds to the model with symmetric $t$-link with 30

TABLE 5
*Tuning parameters for each category in the Reuters dataset chosen by Dataset 1.*

|  | earn | acq | grain | wheat | crude |
|---|---|---|---|---|---|
| $v$ | 8 | 2 | 1 | 1 | 1 |
| $\gamma$ | 0.01 | 0.1 | 0.05 | 0.1 | 0.05 |
| $\delta$ | 1 | 1 | 1.5 | 0.5 | 1.5 |
|  | trade | interest | corn | ship | money-fx |
| $v$ | 1 | 8 | 1 | 2 | 2 |
| $\gamma$ | 0.05 | 0.01 | 0.01 | 0.05 | 0.05 |
| $\delta$ | 1.5 | 1 | 0.5 | 1.5 | 1.5 |

TABLE 6
*Micro $F_1$ and macro $F_1$ measures for different link functions. The last two rows show the average and standard deviation over the five splits.*

| Split | FST | | Probit | | Logistic | |
|---|---|---|---|---|---|---|
|  | micro | macro | micro | macro | micro | macro |
| 1 | 0.862 | 0.801 | 0.855 | 0.781 | 0.857 | 0.786 |
| 2 | 0.853 | 0.796 | 0.849 | 0.788 | 0.853 | 0.795 |
| 3 | 0.863 | 0.802 | 0.859 | 0.788 | 0.861 | 0.793 |
| 4 | 0.871 | 0.807 | 0.864 | 0.788 | 0.866 | 0.795 |
| 5 | 0.874 | 0.816 | 0.867 | 0.802 | 0.869 | 0.804 |
| mean | 0.865 | 0.804 | 0.859 | 0.789 | 0.861 | 0.795 |
| sd | 0.008 | 0.008 | 0.007 | 0.008 | 0.006 | 0.006 |

TABLE 7
*Micro recall and macro recall measures for different link functions. The last two rows show the average and standard deviation over the five splits.*

| Split | FST | | Probit | | Logistic | |
|---|---|---|---|---|---|---|
|  | micro | macro | micro | macro | micro | macro |
| 1 | 0.818 | 0.747 | 0.790 | 0.693 | 0.796 | 0.703 |
| 2 | 0.806 | 0.749 | 0.785 | 0.715 | 0.795 | 0.730 |
| 3 | 0.821 | 0.748 | 0.796 | 0.699 | 0.803 | 0.710 |
| 4 | 0.828 | 0.749 | 0.803 | 0.703 | 0.809 | 0.715 |
| 5 | 0.831 | 0.759 | 0.805 | 0.712 | 0.812 | 0.723 |
| mean | 0.821 | 0.750 | 0.796 | 0.704 | 0.803 | 0.716 |
| sd | 0.01 | 0.005 | 0.008 | 0.009 | 0.008 | 0.011 |

TABLE 8
*Micro precision and macro precision measures for different link functions. The last two rows show the average and standard deviation over the five splits.*

| Split | FST | | Probit | | Logistic | |
|---|---|---|---|---|---|---|
|  | micro | macro | micro | macro | micro | macro |
| 1 | 0.910 | 0.864 | 0.932 | 0.894 | 0.930 | 0.891 |
| 2 | 0.906 | 0.849 | 0.923 | 0.878 | 0.920 | 0.872 |
| 3 | 0.909 | 0.865 | 0.932 | 0.902 | 0.928 | 0.898 |
| 4 | 0.919 | 0.874 | 0.934 | 0.897 | 0.931 | 0.894 |
| 5 | 0.921 | 0.881 | 0.940 | 0.917 | 0.935 | 0.907 |
| mean | 0.913 | 0.867 | 0.932 | 0.898 | 0.929 | 0.892 |
| sd | 0.007 | 0.012 | 0.006 | 0.014 | 0.006 | 0.013 |

degrees of freedom, that approximates the probit link and the third column corresponds to the model with symmetric $t$-link with 8 degrees of freedom. The last two rows of Table 6, 7 and 8 are the average and standard deviation, respectively, accross the five splits of the dataset.

Note that the best performance is achieved by the FST model in the 5 datasets according to the $F_1$ measure and recall. For precision, the best performing model utilizes a probit link.

## 6. Conclusions

This paper introduces a flexible Bayesian generalized linear model for dichotoumous response data.

We gain considerable flexibility by embedding the logistic and probit links into a larger class, the class of all symmetric and asymmetric $t$-link functions. The empirical results and simulations demostrate the good performance of the proposed model. We find that the model with the $t$-link function consistently improves the performance, according to the $F_1$ measure and misclassification rate, as compared with the models with probit or logistic link functions. We compare also our model with the elastic net which is a related method, and showed that our method in general outperforms the elasticnet usually by a substantial margin.

The FST model being a Bayesian model that can also be interpreted as a penalized likelihood model, enjoys all the good properties of these models. Shrinking the parameter estimates for example, is an important property of these models, which has been shown widely that lead to good generalization performance.

We implemented an EM algorithm to learn the parameters of the model. A drawback, is that our algorithm has been implemented to allow only categorical predictors. We plan to extent our work to allow for continuous predictors.

**Acknowledgments.** We are grateful to David D. Lewis for helpful discussions.

## References

[1] ALBERT, J. H. AND CHIB, S. (1993). Bayesian analysis of binary and polychotomous response data. *J. Amer. Statist. Assoc.* **88** 669–679. MR1224394

[2] AZZALINI, A. AND DELLA VALLE, A. (1996). The multivariate skew-normal distribution. *Biometrika* **88** 715–726. MR1440039

[3] CAPOBIANCO, R. (2003). Skewness and fat tails in discrete choice models. MR1986581

[4] CHEN, M., DEY, D. K. AND SHAO, Q. (1999). A new skew link model for dichotomous quantal response data. *J. Amer. Statist. Assoc.* **94** 1172–1186. MR1731481

[5] DEMPSTER, A. P., LAIRD, N. M. AND RUBIN, D. B. (1977). Maximum lilelihood from incomplete data via the EM algorithm (with discussion). *J. Roy. Statist. Soc. Ser. B* **39** 1–38. MR0501537

[6] FERNANDEZ, C. AND STEEL, M. F. J. (1998). On bayesian modelling of fat tails and skewness. *J. Amer. Statist. Assoc.* **93** 359–371. MR1614601

[7] FIGUEIREDO, M. A. T. AND JAIN, A. K. (2001). Bayesian learning of sparse classifiers. *IEEE Computer Society Conference on Computer Vision and Pattern Recognition, Hawaii, December* 2001.

[8] FORMAN, G. (2003). An extensive empirical study of feature selection metrics for text classification. *J. Machine Learning Research*.

[9] GENKIN, A., LEWIS, D. D. AND MADIGAN, D. (2007). Large-scale Bayesian logistic regression for text categorization. *Technometrics*. To appear.

[10] HASTIE, T. J., TIBSHIRANI, R. J. AND FRIEDMAN, J. (2001). *The Elements of Statistical Learning. Data Mining Inference and Prediction.* Springer, New York. MR1851606

[11] MACKAY, D. J. C. (1994). Bayesian non-linear modeling for the energy prediction competition. In *ASHRAE Transactions* **100** 1053–1062.

[12] MLADENIC, D. (1998). Feature subset selection in text-learning. In *European conference on Machine Learning*.

[13] NEAL, R. (1996). *Bayesian Learning for Neural Networks*. Springer, New York.

[14] SCOTT, S. L. (2003). Data augmentation for the Bayesian analysis of multinomial logit models. *Proceedings of the American Statistical Association Section on Bayesian Statistical Science[CD-Rom]*. Amer. Statist. Assoc., Alexandria, VA.

[15] STUKEL, T. A. (1988). Generalized logistic models. *J. Amer. Statist. Assoc.* **83** 426–431. MR0971368

[16] ZOU, H. AND HASTIE, T. (2005). Regularization and variable selection via the elastic net. *J. Roy. Statist. Soc. Ser. B* **67** 301–320. MR2137327